\documentclass[11pt,a4paper]{article}
\usepackage[margin=2.6cm]{geometry}
\usepackage[T1]{fontenc}
\usepackage{lmodern}
\usepackage{amsmath,amssymb,amsthm,mathtools,microtype}
\usepackage{booktabs,array,tabularx,longtable}
\usepackage[shortlabels]{enumitem}
\usepackage{xcolor}
\usepackage{etoolbox}
\usepackage[colorlinks=true,linkcolor=black,citecolor=black,urlcolor=blue!50!black]{hyperref}
\makeatletter
\patchcmd{\LT@output}{\vss}{\vskip\z@\@plus1fil\@minus\normalbaselineskip}{}{}
\patchcmd{\LT@output}{\vss}{\vskip\z@\@plus1fil\@minus\normalbaselineskip}{}{}
\makeatother
\setlist{itemsep=2pt,topsep=4pt}
\theoremstyle{plain}
\newtheorem{theorem}{Theorem}
\newtheorem{lemma}[theorem]{Lemma}
\newtheorem{proposition}[theorem]{Proposition}
\theoremstyle{remark}

\theoremstyle{plain}
\newtheorem{question}[theorem]{Question}
\newcommand{\F}{\mathbb F_2}
\newcommand{\Aff}{\mathrm{Aff}}
\newcommand{\Span}{\operatorname{span}}
\newcommand{\supp}{\operatorname{supp}}
\newcommand{\rad}{\operatorname{rad}}
\newcommand{\rk}{\operatorname{rank}}
\newcommand{\Lam}{\Lambda^2}
\newcommand{\Dir}{E}
\title{The multiplicative complexity of the three-bit low-half multiplier}
\newcommand{\FirstAuthorName}{Bshara Alhodali}
\newcommand{\SecondAuthorName}{Gregory Morse}
\author{\FirstAuthorName\thanks{\raggedright Faculty of Informatics, E\"otv\"os Lor\'and University, Budapest, Hungary.
    \href{mailto:iyzrw6@inf.elte.hu}{\texttt{iyzrw6@inf.elte.hu}},
    \href{mailto:hodalibishara10@gmail.com}{\texttt{hodalibishara10@gmail.com}}.
    ORCID: \href{https://orcid.org/0009-0003-4870-8745}{0009-0003-4870-8745}.}
  \and \SecondAuthorName\thanks{Department of Programming Languages and Compilers,
    Faculty of Informatics, E\"otv\"os Lor\'and University, Budapest, Hungary.
    \href{mailto:morse@inf.elte.hu}{\texttt{morse@inf.elte.hu}}.
    ORCID: \href{https://orcid.org/0000-0002-0231-6557}{0000-0002-0231-6557}.}}
\date{}
\hypersetup{pdftitle={The multiplicative complexity of the three-bit low-half multiplier},
  pdfauthor={\FirstAuthorName\space and \SecondAuthorName}}
\newcommand{\ReleaseCommit}{3770baefb67b5afe1715e22283aa9aed606d096b}
\newcommand{\LeanReleaseStatus}{$\mathrm{MC}(M_3)=6$ is fully formalized in Lean~4.
  The kernel-checked proof uses only the axioms \texttt{propext},
  \texttt{Classical.choice} and \texttt{Quot.sound}.}
\begin{document}
\maketitle
\begin{abstract}
We prove that the multiplicative complexity of the three-bit low-half multiplier is six, giving a structural, enumeration-free lower-bound proof valid against unrestricted nonlinear reuse in Boolean XOR--AND circuits. The value was first established by SAT searches of the first author and independently confirmed by the second author. An explicit six-AND circuit supplies the upper bound. The lower bound combines derivatives, quadratic products, complementary minors, and extraction modulo a Boolean ideal. Independent finite checks accompany the arguments. $\mathrm{MC}(M_3)=6$ is fully formalized in Lean~4, using only Lean's three standard axioms (propositional extensionality, choice and quotient soundness).
\end{abstract}
\setcounter{tocdepth}{1}
\tableofcontents
\clearpage

\section{Introduction}\label{sec:introduction}
Let $a=a_0+2a_1+4a_2$ and $b=b_0+2b_1+4b_2$ be three-bit integers. We study the vectorial Boolean function $M_3(a,b)=(z_0,z_1,z_2)$ specified by $ab\bmod8=z_0+2z_1+4z_2$. Its multiplicative complexity counts two-input AND gates, allowing arbitrary affine combinations of inputs and previously computed gates at no cost. In particular, the factors of an AND gate may contain earlier AND outputs. We use the unrestricted model of Morse~\cite{morse}.

\begin{theorem}\label{thm:MC}
The three-bit low-half multiplier has unrestricted multiplicative complexity
\[
\mathrm{MC}(M_3)=6.
\]
\end{theorem}

An exact value distinguishes an optimal implementation from a merely small one. It measures the number of nonlinear operations required even when linear processing is free, and gives a concrete benchmark for circuit synthesis and lower-bound methods. The carry in $z_2$ makes this a vectorial problem with a quartic output; the degree bound alone does not yield the exact answer.

Boyar, Peralta, and Pochuev~\cite{bpp} study multiplicative complexity over the Boolean basis $(\wedge,\oplus,1)$, and Boyar and Find~\cite{bf} treat vector-valued Boolean functions. Mirwald and Schnorr~\cite{mirwald-schnorr} show that the multiplicative complexity of a quadratic Boolean form is half the rank of the alternating matrix obtained by adding its coefficient matrix to its transpose over $\F$. Morse~\cite{morse} supplies the unrestricted multiplication framework, while Soeken~\cite{soeken} develops SAT methods for determining exact multiplicative complexity. The first author's SAT searches established $\mathrm{MC}(M_3)=6$ by ruling out five AND gates, and the second author confirmed this independently. The accompanying release documents a separate 651-cube LRAT certificate~\cite{cert}, with replayable proofs for zero through four gates and the five-gate cube cover. This computational evidence motivates the following question.

\begin{question}\label{q:structural}
Is there a structural proof of the lower bound $\mathrm{MC}(M_3)\ge6$, one that a reader can verify by hand and that uses no enumeration of circuits and no SAT solving?
\end{question}

Theorem~\ref{thm:MC} answers Question~\ref{q:structural} affirmatively: the lower bound follows from a short list of algebraic lemmas about quadratic Boolean functions and exterior forms (Sections~\ref{sec:toolkit}--\ref{sec:lambda1}), proved against unrestricted nonlinear reuse, and a Lean~4 proof of the same theorem uses the case reductions described in Appendix~\ref{sec:lean}. \c{C}al{\i}k, Turan, and Peralta~\cite{ctp} determine multiplicative complexity for small Boolean functions. Morse~\cite{morse} proves $\mathrm{MC}(\mathrm{Mul}_3)=6$ and $\mathrm{MC}(\mathrm{Mul}_4)=9$ for binary polynomial multiplication in the same unrestricted model; those full polynomial-product maps are distinct from the integer low-half map $M_3$ considered here.

\section{Preliminaries}\label{sec:preliminaries}

Boolean functions on $\F^6$ are identified with their algebraic normal forms in $\F[x]/(x_i^2-x_i)$, and $\Aff=\Span\{1,x_1,\dots,x_6\}$. A circuit with $k$ AND gates is a sequence $g_0,\dots,g_{k-1}$ with
\[
  g_i=L_iR_i,\qquad L_i,R_i\in V_i:=\Aff+\Span\{g_0,\dots,g_{i-1}\},
\]
and it computes a set of targets if all of them lie in $V_k$. This is the unrestricted model of~\cite{morse}. Boyar, Peralta, and Pochuev~\cite{bpp} define multiplicative complexity for sets of Boolean functions by counting the AND gates needed to compute them simultaneously. For any Boolean function $f$, let $\pi_d(f)$ denote its degree-$d$ ANF part. If $\deg f\le2$, write $f=f_0+f_1+f_2$ with $f_d=\pi_d(f)$. Squarefree quadratic monomials are identified with 2-forms, $x_ix_j\leftrightarrow x_i\wedge x_j$.

\paragraph{Targets}
With inputs $a_0,a_1,a_2,b_0,b_1,b_2$ (least significant bit first),
\[
  z_0=a_0b_0,\qquad z_1=a_0b_1+a_1b_0,\qquad z_2=q_2+a_0a_1b_0b_1,\quad q_2=a_0b_2+a_1b_1+a_2b_0 .
\]
The column of weight two of the integer product contains the two bits $a_0b_1$
and $a_1b_0$. Its carry into the column of weight four is their product, which
gives the quartic term of $z_2$. The column of weight one contains a single
bit and sends no carry.
We write
\[
  \Omega=a_0\wedge a_1\wedge b_0\wedge b_1,\qquad P_0=a_0\wedge b_0,\qquad Z_1=a_0\wedge b_1+a_1\wedge b_0 .
\]

\paragraph{Linear gates}
A gate $g_i=L_iR_i$ is \emph{linear} if $L_i$ and $R_i$ are \emph{equal as functions} to affine forms, even if they are written using earlier gates. This is a semantic notion.

\begin{lemma}[Normal form]\label{l4:lem:normal}
If a $k$-gate circuit computes $(z_0,z_1,z_2)$, so does a $k$-gate circuit in which $g_0=a_0b_0$, gates $0,\dots,\lambda-1$ are linear and gates $\lambda,\dots,k-1$ are not, for some $\lambda\ge1$.
The normalization preserves the number of linear gates unless the replaced gate is nonlinear, in which case it increases that number by one.
\end{lemma}

\begin{proof}
Let $j$ be the first index with $z_0\in V_{j+1}$. Then $g_j=z_0+v$ for some $v\in V_j$. Replace $g_j$ in place by the direct product $a_0b_0$. The wire space immediately after this gate is the same, so every later gate can keep its factor functions. Move this direct gate to the first position by adjacent swaps. In such a swap, the later direct gate needs only $\Aff$, and the displaced gate retains its old factors. The two gates span the same extension of the preceding space. Repeating these swaps moves all remaining linear gates immediately after $g_0$, in their original order. The number of linear gates increases by one if the replaced gate was nonlinear, and is preserved otherwise.
\end{proof}

For any space $Q$ of 2-forms, put
\[
 V(Q)=\{f:\deg f\le2,\ \pi_2(f)\in Q\}.
\]
\begin{lemma}\label{l4:lem:Qonly}
If the first $j$ gates have affine factors and their quadratic parts span $Q$, then $V_j=V(Q)$.
\end{lemma}

\begin{proof}
If $\pi_2(f)=\sum_i\alpha_iP_i$, then $f-\sum_i\alpha_ig_i$ has no quadratic part and therefore lies in $\Aff$. The other inclusion is clear.
\end{proof}

Let $U=\langle a_0,a_1,b_0,b_1\rangle$, so $\Lam(U)$ is the space of 2-forms with no monomial involving $a_2$ or $b_2$. The \emph{support} $\supp\omega$ of a 2-form $\omega$ is the smallest subspace $W$ of linear forms with $\omega\in\Lam(W)$, equivalently the row space of its alternating matrix; $\dim\supp\omega$ is the rank of $\omega$. Every 2-form decomposes uniquely as
\[
  \omega=\omega_U+a_2\wedge\alpha+b_2\wedge\beta+\gamma\,a_2\wedge b_2,\qquad \omega_U\in\Lam(U),\ \alpha,\beta\in U,\ \gamma\in\F,
\]
and we define the \emph{cross part} $\Phi(\omega)=(\alpha,\beta,\gamma)$; equivalently, $\Phi$ records the contractions of $\omega$ with the unit vectors $e_{a_2},e_{b_2}\in\F^6$. Note $\Phi(q_2)=(b_0,a_0,0)$.

\paragraph{Directions} Let $\Dir=\{e_{a_2},\,e_{b_2},\,e_{a_2}+e_{b_2}\}\subset\F^6$, a set of three vectors. For any direction $e$, $\iota_e$ is contraction with $e$, and $l(e)$ denotes the value of the linear form $l$ at $e$.

The radical $\rad\omega$ is the kernel of the alternating matrix of $\omega$, a space of direction vectors; thus $\supp\omega=(\rad\omega)^\circ$. We sometimes abbreviate the directions $e_{a_2}$, $e_{b_2}$, $e_{a_2}+e_{b_2}$ by $a_2$, $b_2$, $a_2+b_2$ when they occur as subscripts of contractions or derivatives. A direction $e\in E$ \emph{kills} $Q$ if $\iota_e(Q)=0$.
For $e\in E$, the difference $D_ez_2$ equals $\delta_e=\iota_e(q_2)$, respectively $b_0$, $a_0$, or $a_0+b_0$.
When a 2-form occurs in a Boolean expression, it denotes its homogeneous quadratic ANF representative. Likewise the carry function $a_0a_1b_0b_1$ is occasionally denoted by $\Omega$ in Boolean identities; exterior products use the corresponding 4-form.

Let $A$ be a space of Boolean functions and let $f,h\in A$.
We say that the product $fh$ \emph{places} $T$ from $A$ if
$T\in A+\langle fh\rangle$ and $T\notin A$.
Over $\F$, if $T\notin A$, this is equivalent to $fh+T\in A$.

We use $V_4(Q)$ as a synonym for $V(Q)$ in the four-linear-gate statements. Their single-product feasibility condition is
\begin{equation}\label{l4:eq:star}
 f,h\in V_4(Q),\qquad fh\in z_2+V_4(Q).
\end{equation}

\begin{lemma}[Support]\label{l4:lem:support}
If a rank-4 form $Z$ equals $P+P'$ with $P,P'$ of rank 2, then $P,P'\in\Lam(\supp Z)$. In particular every splitting of $Z_1$ or $Z_1+P_0$ into two rank-2 forms lies in $\Lam(U)$.
\end{lemma}

\begin{proof}
$\supp Z\subseteq\supp P+\supp P'$, and both sides have dimension $4$, so they are equal. The second claim follows since $Z_1$ and $Z_1+P_0=a_0\wedge(b_0+b_1)+a_1\wedge b_0$ both have rank $4$ and support $U$. Lean checks the two target splittings by finite computation in \nolinkurl{lemma3_finite_Z1} and \nolinkurl{lemma3_finite_Z1P0}.
\end{proof}

\paragraph{Cross-part kernel}\label{l4:lem:cross} The kernel of $\Phi$ is $\Lambda^2(U)$, since $\Phi$ records exactly the coefficients involving $a_2$ or $b_2$.

\begin{lemma}\label{l4:lem:noq2}
No wedge $R=l\wedge r$ satisfies $\Phi(R)=\Phi(q_2)$.
\end{lemma}

\begin{proof}
Both contractions $\iota_{e_{a_2}}R,\iota_{e_{b_2}}R$ lie in $\langle l,r\rangle$. If they were $b_0$ and $a_0$, then $\langle l,r\rangle=\langle a_0,b_0\rangle$ and $R=a_0\wedge b_0$, whose $a_2$-contraction is $0$. Lean checks this step by finite computation in \nolinkurl{lemma6}.
\end{proof}

For a linear form $x$ let $\pi(x)=(x(e_{a_2}),x(e_{b_2}))\in\F^2$, its pair of coefficients of $a_2$ and $b_2$. For $v\in\F^2\setminus\{0\}$ and $r\in U$ write $v\otimes r:=(v_1r,\,v_2r,\,0)$.

\begin{lemma}[Types]\label{l4:lem:types}
Every wedge has a representation $R=l\wedge r$ with one of the following factor types.
\begin{enumerate}[(a)]
  \item \emph{Core:} $l,r\in U$, so $\Phi(R)=0$.
  \item \emph{Type $v$:} one can choose $r\in U$ and $\pi(l)=v\neq0$. Then $\Phi(R)=v\otimes r$ and $\iota_eR=l(e)\,r$ for $e\in\Dir$; in particular $\iota_eR=0$ for the unique $e\in\Dir$ with $l(e)=0$, namely $e=e_{b_2},\ e_{a_2},\ e_{a_2}+e_{b_2}$ for $v=(1,0),(0,1),(1,1)$.
  \item \emph{Full:} $R=(a_2+x)\wedge(b_2+y)$ with $x,y\in U$, and $\Phi(R)=(y,x,1)$.
\end{enumerate}
\end{lemma}

\begin{proof}
Use two zero forms for a zero wedge. Otherwise classify the dimension of the image $\pi(\langle l,r\rangle)$. In dimension zero both factors are already in $U$. If it is a line $\langle v\rangle$, choose the nonzero $r$ in the kernel of $\pi$ restricted to the factor plane, and choose $l$ with $\pi(l)=v$. If it is all of $\F^2$, choose $l,r$ with $\pi(l)=(1,0)$ and $\pi(r)=(0,1)$. The formulas follow by contraction. Lean checks the two-bit change of basis in \nolinkurl{types_finite}.
\end{proof}

Among the 651 rank-2 forms on $\F^6$, 35 lie in $\Lam(U)$, 360 have type $v$ and 256 are full.

\section{A six-gate upper bound}\label{sec:upper}
Use the six AND gates
\begin{align*}
g_0&=a_0b_0,&g_1&=a_0b_1,&g_2&=a_1b_0,\\
g_3&=a_0b_2,&g_4&=a_2b_0,&g_5&=(g_1+a_1)(g_2+b_1).
\end{align*}
The outputs are
\[
 z_0=g_0,\qquad z_1=g_1+g_2,\qquad
 z_2=g_5+g_1+g_2+g_3+g_4.
\]
Indeed $g_5=a_0a_1b_0b_1+g_1+g_2+a_1b_1$, so the stated output is exactly the target ANF. Hence $\mathrm{MC}(M_3)\le6$.

\section{Lower-bound overview}\label{sec:overview}
By Lemma~\ref{l4:lem:normal}, a five-gate circuit may be assumed to start with $g_0=a_0b_0$ and to have $\lambda$ initial linear gates, followed by nonlinear gates, for some $1\le\lambda\le5$. Linearity is semantic: it concerns the two factor functions.
The quadratic spans may have redundant generators.

Degree excludes $\lambda=5$. For $\lambda=4$ a good external direction excludes the final product. The $\lambda=3$ proof uses the same direction argument or a $U$-direction with a nonzero middle-bit component. The smaller prefixes use the coset obstruction, multiplier rigidity and extraction modulo the low-bit ideal. The quadratic first gate at $\lambda=1$ uses the same three-product tail argument as $\lambda=2$.

\section{Five linear gates}\label{sec:lambda5}
\begin{theorem}\label{thm:lambda5}
No five-AND circuit with five linear gates computes $(z_0,z_1,z_2)$.
\end{theorem}
\begin{proof}
Each gate is a product of two affine functions, so every function in $V_5$ has degree at most two. The target $z_2$ has the nonzero quartic part $a_0a_1b_0b_1$.
\end{proof}

\section{Toolkit}\label{sec:toolkit}

\subsection{Directions and affine hyperplanes}

For a direction vector $e$, write $D_e p(x)=p(x+e)+p(x)$. If $p$ has degree at most two, expansion of each quadratic monomial gives
\[
D_e p(x)=D_e p(0)+(\iota_e\pi_2p)(x).
\]
In particular, $D_dD_e p$ is the constant $(\iota_e\pi_2p)(d)$.
For arbitrary Boolean functions,
\[
D_e(fh)=(D_ef)h+(D_eh)f+(D_ef)(D_eh).
\]
We call $e\in E$ a \emph{good direction} for $Q$ if
$\iota_e(Q)\subseteq\langle\ell\rangle$ for some linear form
$\ell\ne\delta_e$. The same direction works for every subspace of $Q$ and for $Q+\langle Z_1\rangle$, since $\iota_e Z_1=0$.

An affine function vanishing on a nonempty affine hyperplane has linear part in the span of its normal. Indeed its linear part kills every difference of two points on the hyperplane. This observation also applies to the whole space, with normal zero.

\begin{lemma}\label{toolkit:common}
If each generator of $Q$ has cross part zero or $\Phi(a\wedge b)$ for fixed linear forms $a,b$, then $Q$ has a good direction.
\end{lemma}
\begin{proof}
By Lemma~\ref{l4:lem:noq2}, $\Phi(a\wedge b)\ne\Phi(q_2)$, so for at least one of $e=e_{a_2},e_{b_2}$ the contraction $\ell=\iota_e(a\wedge b)$ differs from $\delta_e$. By linearity every contraction of $Q$ along $e$ is in $\langle\ell\rangle$.
\end{proof}

\begin{lemma}[Derivative lemma]\label{l4:lem:deriv}
If $Q$ has a good direction, there are no $f,h\in V(Q)$ with $fh+z_2\in V(Q)$.
\end{lemma}
\begin{proof}
Write $g=fh+z_2$. The three derivatives along a good direction have the form $c+\gamma\ell$. Swap the factors if necessary, or replace $f$ by $f+h$ if both have nonzero line coefficient. In the latter case the product changes by $h^2=h$, so the placement persists. We can thus assume $D_e f=c$ is constant. Put $H=D_e h$ and $A=D_e g+\delta_e$. The linear part of $A$ is outside $\langle\ell\rangle$, because $\delta_e$ is nonzero and differs from $\ell$. The product rule gives
\[
ch+Hf+cH=A.
\]
If $c=1$, then $h=H(f+1)+A$, whence $fh=fA$. If $c=0$, then $Hf=A$. Were $H\ne1$, its zero set would be a nonempty level set with normal in $\langle\ell\rangle$. The affine function $A$ would vanish there, which its linear part precludes. Thus $H=1$ and $f=A$. In both branches the product has degree at most three, contrary to its carry coefficient in $fh=z_2+g$. Lean checks the affine zero-set step by finite computation in \nolinkurl{line_finite_b0}, \nolinkurl{line_finite_a0}, and \nolinkurl{line_finite_a0b0}.
\end{proof}

\begin{lemma}[W]\label{toolkit:w}
Let $\deg G\le2$ and let $a$ be a nonzero linear form. If $G$ vanishes on $a(x)=\alpha$, then $\pi_2G=a\wedge m$ for some linear form $m$.
\end{lemma}
\begin{proof}
Write $\omega=\pi_2G$. For $e\in\ker a$, $D_eG$ also vanishes on the hyperplane. Thus $\iota_e\omega$ lies in $\langle a\rangle$. Choose $v$ with $a(v)=1$ and put $m=\iota_v\omega$. For $e\in\ker a$ the coefficient of $a$ in $\iota_e\omega$ is $\omega(e,v)=m(e)$. Contraction of $a\wedge m$ agrees with that of $\omega$ on $\ker a$ and on $v$, since $m(v)=0$. These directions span the space.
\end{proof}

\begin{lemma}[Feedback normalization]\label{toolkit:feedback}
Let $A$ be a space of Boolean functions containing $1$ and $W=A+\langle g\rangle$. If $F,H\in W$ and $FH+T\in W$, then either some $f,h\in A$ satisfy $fh+T\in W$, or some $k,C\in A$ satisfy $k(g+C)+T\in A$.
\end{lemma}
\begin{proof}
Write $F=f+sg$, $H=h+tg$, $FH+T=v+\sigma g$, with $f,h,v\in A$.
If $s=t=0$, the first conclusion holds. With exactly one $g$ coefficient nonzero, swap to obtain $F=f$, $H=h+g$, and use $k=f+\sigma$, $C=h$. With both coefficients nonzero,
$(f+g)(h+g)=(f+h+1)(g+f)$ by idempotence, so use $k=f+h+1+\sigma$, $C=f$. In either feedback case $k(g+C)+T=v+\sigma C$.
\end{proof}

\subsection{One general coset obstruction}

For a rank-two $2\times4$ matrix $M$ over $\F$, the minors of a kernel-basis matrix $N$ are the complementary minors of $M$. Invertible row reduction and a column permutation put $M=[I_2\mid T]$, with $T=\left(\begin{smallmatrix}a&b\\c&d\end{smallmatrix}\right)$. Row operations preserve all two-column determinants over $\F$, since every invertible determinant is one. A kernel basis is
\[
N=[T^T\mid I_2]=\begin{pmatrix}a&c&1&0\\b&d&0&1\end{pmatrix}.
\]
The minors of $M$, in the order $12,13,14,23,24,34$, are
\[
(1,c,d,a,b,ad+bc),
\]
and the minors of $N$ in that order are
\[
(ad+bc,b,a,d,c,1).
\]
Thus $\det_{jk}N=\det_{lm}M$ for complementary pairs $\{j,k\},\{l,m\}$. Changing the kernel basis also has determinant one, and undoing the column permutation preserves these equalities in characteristic two.

\begin{lemma}[Gram identity]\label{toolkit:gram}
Let $M$ have two independent rows over $\F$ and four columns, and suppose $\det_{12}M=\det_{34}M$. For every symmetric $4\times4$ matrix $G$ with zero diagonal and $MG=0$, one has $G_{12}=G_{34}$.
\end{lemma}
\begin{proof}
Let $N$ have a basis of $\ker M$ as its rows. Since every column of $G$ is in $\ker M$ and $G$ is symmetric, $G=N^T C N$ for an alternating $2\times2$ matrix $C$. To see this, choose a right inverse $K$ of $N$. Then $N^TK^T$ fixes the columns of $G$, so by symmetry $G=N^T(K^TGK)N$, with $K^TGK$ alternating. Hence $G_{ij}=C_{12}\det_{ij}N$. The complementary-minor identity above gives the result. Lean checks the corresponding Gram step by finite computation in \nolinkurl{gram_finite} over $2^{14}$ assignments.
\end{proof}

For core functions $X,Y$ put $\beta(X,Y)=[a_0a_1b_0b_1](XY)$. In particular, for quadratic functions this is the exterior carry pairing of their quadratic parts.

\begin{lemma}[General coset lemma, including cubic cores]\label{toolkit:coset}
Let $\mathcal B$ be a vector space of functions of $u=(a_0,a_1,b_0,b_1)$ containing $\Aff(U)$. For the pairing $\beta(X,Y)=\sum_{u\in\F^4}X(u)Y(u)$, assume $1,a_0,b_0\in\rad\beta$. Put
\[
\mathcal A=\mathcal B\oplus\langle a_2,b_2\rangle,\qquad f,h\in\mathcal A,\qquad g=fh,\qquad\mathcal W=\mathcal A+\langle g\rangle.
\]
Then:
\begin{enumerate}[(a)]
\item no $F,H\in\mathcal A$ satisfy $FH+z_2\in\mathcal W$;
\item no $k,F'\in\mathcal A$ satisfy $k(F'+g)\in z_2+\mathcal A$.
\end{enumerate}
Consequently $z_2\notin\mathcal W$, and no product of two elements of $\mathcal W$ belongs to $z_2+\mathcal W$.
In particular this holds for $\mathcal A=V(Q)$ for every $Q\subseteq\Lambda^2(U)$, for every dimension of $Q$ and without rank-two generation, and every product $g=fh$ with $f,h\in V(Q)$.
\end{lemma}

\begin{proof}
An element of $\mathcal B$ has zero quartic coefficient since its parity is $\beta(1,X)=0$. Also $\beta(X,X)=\beta(1,X)=0$ by idempotence, so $\beta$ is alternating. This is the abstract coset-base hypothesis: the two external differences of $\mathcal A$ are constant, and conversely that condition gives the displayed decomposition after subtracting external linear terms. Write $p=(p_1,p_2)=(a_2,b_2)$, and
\[
z_2=\kappa+p_1b_0+p_2a_0,\qquad \kappa=a_1b_1+a_0a_1b_0b_1.
\]
Every element of $\mathcal A$ has a unique core part in $\mathcal B$ and a scalar external linear part. Write
\[
f=\varphi+c(p),\quad h=\psi+d(p),\quad
F=X+\ell(p),\quad H=Y+m(p),\quad
v=C+n(p)+\sigma g,
\]
with core parts in $\mathcal B$, external linear forms $c,d,\ell,m,n$, and $\sigma\in\F$.

\emph{Part (a): coefficient comparison.} Suppose that $FH+z_2+v=0$. In particular this sum is independent of the external coordinates and has zero carry coefficient. Set $\lambda_1=b_0$, $\lambda_2=a_0$. Comparing the coefficients of $p_i$ modulo constant core functions and then of $p_1p_2$ gives
\begin{align}
\ell_iY+m_iX+\sigma c_i\psi+\sigma d_i\varphi&\equiv\lambda_i\pmod{\F},\quad i=1,2,\label{coset:linear}\\
\det(\ell,m)+\sigma\det(c,d)&=0.\label{coset:quadratic}
\end{align}
Terms such as $\ell_i m_i p_i^2$ are constant multiples of $p_i$ and are included in the discarded constants in \eqref{coset:linear}. Lean checks the linear comparison in \nolinkurl{fin_E1}. At $p=0$, the carry coefficient is
\begin{equation}\label{coset:carry}
\beta(X,Y)+\sigma\beta(\varphi,\psi)=1,
\end{equation}
because $C$ has no quartic coefficient and $\kappa$ has coefficient one.
Put $x=(Y,X,\psi,\varphi)$ and let $M$ be the $2\times4$ matrix with columns $\ell,m,\sigma c,\sigma d$. Its rows are independent: a nonzero row relation in \eqref{coset:linear} would make a nonzero linear combination of $a_0,b_0$ constant. Thus $\rk M=2$.

Put $G_{jk}=\beta(x_j,x_k)$. It is alternating. Pairing the two equations in \eqref{coset:linear} with each $x_k$ gives $MG=0$, because $1,a_0,b_0$ are in the radical. Equation \eqref{coset:quadratic} says the complementary minors $\det_{12}M$ and $\det_{34}M$ are equal. Lemma~\ref{toolkit:gram} yields $G_{12}=G_{34}$. If $\sigma=1$, this contradicts \eqref{coset:carry}. If $\sigma=0$, the first two columns of $M$ form an invertible block and its other two columns are zero. Hence $MG=0$ forces the first two rows of $G$ to vanish. Then $G_{12}=0$, again contradicting \eqref{coset:carry}. This proves (a).

\emph{Part (b) via external differences.} Suppose $k(F'+g)=z_2+v$ with $k,F',v\in\mathcal A$, and put $G=F'+g$. For each nonzero external direction $e\in E$, let
\[
D_ek=\mu_e,\quad D_eF'=a_e,\quad D_ev=v_e,\qquad
L_e=D_eg=c_eh+d_ef+c_ed_e\in\mathcal A.
\]
All of $\mu_e,a_e,v_e,c_e,d_e$ are scalars. Differentiation of $kG=z_2+v$ gives the exact identity
\begin{equation}\label{coset:feedback}
\mu_eG+(k+\mu_e)(a_e+L_e)=\delta_e+v_e.
\end{equation}
If some $\mu_e=1$, solve for $G$ and multiply by $k$. The Boolean identity $k(k+1)=0$ gives
\[
kG=k(\delta_e+v_e)=z_2+v.
\]
Both factors on the middle expression belong to $\mathcal A$, contradicting (a) with an output correction in $\mathcal A$.

Otherwise $k=\zeta(u)\in\mathcal B$. It is nonzero, since $z_2\notin\mathcal A$ by its carry coefficient. In \eqref{coset:feedback} the only external dependence on the left is
\[
\zeta(u)\bigl(c(e)d(p)+d(e)c(p)\bigr).
\]
A nonzero function of $u$ times a scalar external linear form vanishes identically only if that linear form is zero: evaluate at a core input with $\zeta=1$. Hence $c(e)d+d(e)c=0$ for both external basis directions. Equivalently $\det(c,d)=0$, so $c,d$ are dependent.

Since $c,d$ are dependent, either $e_{a_2}$ kills both, $e_{b_2}$ kills both, or the two nonzero coefficient vectors $(c_1,d_1),(c_2,d_2)$ agree and $e_{a_2}+e_{b_2}$ kills both. Choose a killing direction $e_0$ from these three cases. Equation \eqref{coset:feedback} gives $a_{e_0}=1$ and $k=\delta_{e_0}+v_{e_0}$. Choose an external basis direction $e_1\ne e_0$. The two low-bit forms $\delta_{e_0},\delta_{e_1}$ are independent, so there is a point where $\delta_{e_0}=v_{e_0}$ and $\delta_{e_1}=1+v_{e_1}$. Equation \eqref{coset:feedback} in direction $e_1$ then has left side zero and right side one. Lean checks the scalar steps in \nolinkurl{fin_mu1}, \nolinkurl{fin_mu0}, \nolinkurl{fin_D}, \nolinkurl{fin_onept}, and \nolinkurl{fin_case3}.

\emph{Final product and applications.} Taking $F=H=0$ in (a) gives $z_2\notin\mathcal W$. For a proposed product from $\mathcal W$, apply Lemma~\ref{toolkit:feedback} with $T=z_2$. Its ordinary conclusion is excluded by (a) and its feedback conclusion by (b).
For $Q\subseteq\Lambda^2(U)$ choose $\mathcal B=\{X(u):\deg X\le2,\ X_2\in Q\}$. Products of $1,a_0,b_0$ with its elements have degree at most three, so all three are in $\rad\beta$. No step of the proof used $\dim Q$ or rank-two generators. For a cubic core the hypotheses stated in the lemma supply exactly the same space $\mathcal B$ and the proof above already applies to it.
\end{proof}

\subsection{Quadratic extraction modulo the low-bit ideal}
Let $I=(a_0b_0)$ be the ideal of functions vanishing on the three low-bit fibers $(a_0,b_0)\ne(1,1)$. Congruence modulo $I$ is equality on these fibers.

\begin{lemma}[Quadratic extraction]\label{l2:lem:ideal}
Let $Q=\langle P_0,l\wedge r\rangle$, with $l,r$ linear. If $f,h\in V(Q)$ and $R$ has degree at most two with $fh\equiv R\pmod I$, then $\pi_2R=\pi+a\wedge b$ for some $\pi\in Q$ and linear $a,b$.
\end{lemma}
\begin{proof}
Modulo $I$, each factor is affine plus a scalar multiple of $lr$. If one is affine modulo $I$, take it as the second factor $B$, with first factor $F\in V(Q)$. Otherwise replace $h$ by $f+h$, which is affine modulo $I$, and $R$ by $R+f$, since $f(f+h)=fh+f$. This changes the resulting quadratic part only by an element of $Q$. It suffices to handle $FB\equiv R\pmod I$ with $B$ affine.

Write $B=B(0)+a$. On $a(x)=\alpha$, the quadratic $G=R+(B(0)+\alpha)F$ vanishes off the $(1,1)$ fiber. If $a\ne0$ involves only $a_0,b_0$, take $\alpha=a_{a_0}+a_{b_0}+1$. Then the entire level set avoids $(1,1)$ and Lemma~\ref{toolkit:w} applies to $G$.

Otherwise take $\alpha=0$. There are directions $e,d\in\ker a$ whose low-bit projections are $(1,0),(0,1)$. For $a=0$ use the low-bit unit directions. If $a_j=1$ at a free coordinate, add $a_{a_0}e_j$ and $a_{b_0}e_j$ to the respective low-bit unit directions. For any point of the level set with low bits $(1,1)$, the other three corners of its $e,d$ square are off that fiber. Thus $G(x)=D_dD_eG(x)=c$, a constant by the quadratic derivative formula. On the entire level set $G=ca_0b_0$. Subtract this correction, which belongs to $V(Q)$. If $a\ne0$, Lemma~\ref{toolkit:w} extracts the wedge from the corrected function. If $a=0$, that function is zero everywhere and the wedge is zero.
\end{proof}
The same square with $a=0$ shows that every quadratic function in $I$ is a scalar multiple of $a_0b_0$.

\begin{lemma}\label{l2:lem:rankext}
For all $\varepsilon,\alpha\in\F$ and linear forms $l,r,a,b$,
\[
q_2+\varepsilon Z_1+\alpha P_0+l\wedge r\ne a\wedge b .
\]
\end{lemma}
\begin{proof}
The four forms have a common nonzero zero $e$, since their common kernel has dimension at least two. Then $\iota_e$ kills both wedges. Yet the contraction of $q_2+\varepsilon Z_1+\alpha P_0$ has external coordinates $e_{b_0},e_{a_0}$. If it is zero, these low coordinates of $e$ vanish. Its middle coordinates then give $e_{b_1}=e_{a_1}=0$, and its low coordinates give $e_{b_2}=e_{a_2}=0$, a contradiction. Lean checks the radical computation in \nolinkurl{q2_iota_ne_zero}.
\end{proof}

\begin{lemma}[Feedback exclusion modulo $I$]\label{toolkit:modfeedback}
Let $Q=\langle P_0,l\wedge r\rangle$ and $S=Q+\langle Z_1\rangle$. If $f,h\in V(Q)$ and $F,v\in V(S)$, then $fh\not\equiv q_2+F+v\pmod I$.
\end{lemma}
\begin{proof}
An equality would give, by Lemma~\ref{l2:lem:ideal}, a wedge in $q_2+S$. The coefficient of $l\wedge r$ may be zero, in which case use the zero wedge in Lemma~\ref{l2:lem:rankext}. Both values are excluded.
\end{proof}

\subsection{Two ordinary products with a cubic signal}
We call a Boolean function core supported when all its terms of degree at least two use only coordinates in $U$.

\begin{lemma}[Two ordinary products cannot place $z_2$]\label{l1:lem:ordinarypair}
For homogeneous $l,B$ with $B\notin\langle a_0,b_0\rangle$, set $c=(P_0+l)B$, $A_0=V(\langle P_0,Z_1\rangle)$ and $A=A_0+\langle c\rangle$. For all $F_1,F_2,F_3,F_4\in A$ one has $F_1F_2+F_3F_4+z_2\notin A$.
\end{lemma}
\begin{proof}
Suppose $c$ is core supported. Set $c_U=c|_{a_2=b_2=0}$.
Then $c+c_U\in\langle a_2,b_2\rangle$ and
$A=\mathcal B\oplus\langle a_2,b_2\rangle$, where
$\mathcal B=\Aff(U)+\langle P_0,Z_1,c_U\rangle$.
Each of $X,a_0X,b_0X$ has degree at most three for $X\in\mathcal B$.
Indeed, all cubic terms of $c_U$ contain both low bits, and
multiplication by either of these bits preserves those terms.
All three functions therefore have even parity on $\F^4$, proving
$1,a_0,b_0\in\rad\beta$. Apply Lemma~\ref{toolkit:coset}
with $g=F_1F_2$. The alleged identity would give
$F_3F_4+z_2\in A+\langle g\rangle$, a contradiction.

Now assume $c$ is noncore and write $L=l$.

\emph{Step 1. Compare coefficients on two hyperplanes.}
Write $F_i=A_i+t_ic$ with $A_i\in A_0$. Expanding and using $c^2=c$ puts the asserted identity in the form
\[
cH=z_2+v_0+A_1A_2+A_3A_4,\qquad H,v_0\in A_0.
\]
Products of two elements of $A_0$ have degree at most three, since all wedges in $\langle P_0,Z_1\rangle$ vanish. Thus $cH$ has quartic part exactly $\Omega$. Since $B(cH)=cH$, its degree-five part gives $B\wedge\Omega=0$, so $B\in U$. Cubicity gives $B\notin\langle a_0,b_0\rangle$. Noncore support then forces $L$ to have nonzero external part. Choose external coordinates $p,q$ with
\[
L=L_U+p,\qquad c=(P_0+L_U)B+pB,
\qquad (q_2)_{\rm ext}=sp+rq,
\]
where $r,s$ are independent low-bit linear forms. Write $A_i$ with external part $c_i p+d_iq$, and core part an affine function plus $w_iP_0+z_iZ_1$. The symbols $t,w,z,c_*,d$ denote the resulting scalar four-vectors. On $B=j$, for $j=0,1$,
\[
F_i=x_i(j)+\alpha_i(j)p+d_iq,\quad
\alpha(j)=c_*+jt,\quad
x_i(j)=(A_i)_U+jt_i(P_0+L_U),\quad w(j)=w+jt.
\]
The coefficients of $x(j)$ in the independent degree-two parts of $P_0\vert_{B=j}$ and $Z_1\vert_{B=j}$ are $w(j),z$, as proved below.

Let $J$ interchange coordinates 1 and 2, and 3 and 4; its associated alternating pairing is $v^TJw$. Let the two rows of $M(j)$ be $(J\alpha(j))^T$ and $(Jd)^T$. The external linear coefficients of the proposed identity, modulo constant functions on $B=j$, and its $pq$ coefficient are
\begin{equation}\label{parity:coefficients}
M(j)x(j)\equiv(s,r)^T\pmod\F,\qquad
\alpha(j)^TJd=0.
\end{equation}
The output correction from $A$ has only scalar external coefficients on $B=j$; hence it is absorbed in the constants here. The restrictions of $r,s$ remain independent modulo constants: a nonzero low-bit linear combination could be constant on $B=j$ only if it equaled $B$ up to a constant, contrary to $B\notin\langle a_0,b_0\rangle$. Thus $\rk M(j)=2$.
The restrictions of $P_0,Z_1$ also have independent quadratic parts. If a quadratic restricts to an affine function on $B=j$, subtract an affine extension of that restriction. Lemma~\ref{toolkit:w} then expresses its quadratic part as $B\wedge m$, of rank at most two and with $B$ in its support when nonzero. This is impossible for $P_0$, since $B\notin\supp P_0$, and for $Z_1$ or $Z_1+P_0$, since they have rank four. Comparing quadratic parts of \eqref{parity:coefficients} therefore gives
\[
M(j)w(j)=M(j)z=0.
\]

\emph{Step 2. Apply the Gram identity to the coefficient vectors.}
The two-row matrix $M=M(j)$ has equal complementary minors $\det_{12}M=\det_{34}M$, because their sum is $\alpha(j)^TJd=0$. Apply Lemma~\ref{toolkit:gram} to $G=w(j)z^T+zw(j)^T$. This matrix is alternating and $MG=0$ since $M$ kills $w(j),z$. Thus $w(j)^TJz=G_{12}+G_{34}=0$ for both $j$. It follows that $w^TJz=t^TJz=0$.

\emph{Step 3. Apply the Gram identity to the slice pairing.}
On $B=j$, write $\beta_j(X,Y)=\sum_{u:B(u)=j}X(u)Y(u)$ and put $y_i=x_i+z_iZ_1$. These functions belong to the restriction of $T=\Aff(U)+\langle P_0\rangle$. The pairing on $T$ is alternating, since its functions are quadratic on a three-bit affine space and therefore have zero parity. Its radical contains $1,r,s$: products of $P_0$ with $r,s\in\langle a_0,b_0\rangle$ have degree at most two by Boolean idempotence. The products of affine functions also have degree at most two. By $Mz=0$ and \eqref{parity:coefficients}, the two linear combinations $My$ lie in $\langle 1,r,s\rangle$. Thus $G_{ik}=\beta_j(y_i,y_k)$ is alternating and $MG=0$. Lemma~\ref{toolkit:gram} gives $\beta_j(y_1,y_2)+\beta_j(y_3,y_4)=0$.

\emph{Step 4. Sum over $j=0,1$.}
Put $D_j=\sum_i(Jz)_ix_i(j)$. Its $j$-dependent part is $j(t^TJz)(P_0+L_U)=0$, and its coefficients of $P_0,Z_1$ are $w^TJz=0$ and $z^TJz=0$. Hence it is the restriction of a single affine core function $D$, independent of $j$.
The parity pairing is alternating on the larger space of all restricted quadratic functions as well. Expanding $y_i=x_i+z_iZ_1$ and using Step 3 therefore gives
\[
\beta_j(x_1,x_2)+\beta_j(x_3,x_4)=\beta_j(Z_1,D).
\]
The possible diagonal term is a scalar multiple of $\beta_j(Z_1,Z_1)=\sum_{B=j}Z_1=0$. Adding the two hyperplanes now yields
\[
\sum_{j=0}^1\bigl(\beta_j(x_1,x_2)+\beta_j(x_3,x_4)\bigr)
=\sum_{u\in\F^4}Z_1(u)D(u)=0,
\]
since $Z_1D$ has degree at most three on four bits.

\emph{Step 5. Compare with the carry parity.}
At $p=q=0$ the proposed identity equates $x_1x_2+x_3x_4$, on each hyperplane, to $a_1b_1+\Omega$ plus the core restriction of an element of $A$. This correction has degree at most three. A monomial on four bits has even parity unless it is the full monomial, whose parity is one: summing a degree-$d$ monomial gives $2^{4-d}$ ones. The total parity of the right side is therefore one, contradicting Step 4.

\end{proof}

\subsection{Last-factor rigidity}

Put $A_0=V(\langle P_0,Z_1\rangle)$, and for homogeneous linear forms $l,b$ and $d\in\F$ set $c=(dP_0+l)b$. Write $A=A_0+\langle c\rangle$.

\begin{lemma}[Last-factor rigidity modulo $I$]\label{l2:lem:lastfactor}
Suppose $b\notin\langle a_0,b_0\rangle$ and $c$ is noncore. If $k\in A$, $R+z_2\in A$ and $kR=R$, then $k\equiv1\pmod I$.
\end{lemma}
\begin{proof}
Write $k=u+\tau c$, $R=z_2+v+\sigma c$ with $u,v\in A_0$. Along any external direction $e$, $D_eR$ belongs to $A_0$. Indeed $D_ez_2$ is linear, the difference of $v$ is constant, and
\[
D_e c=b(e)(dP_0+l)+l(e)b+l(e)b(e).
\]
If $D_e k=1$, the equations $kR=R$ at $x$ and $x+e$ give $R=kD_eR$: both sides vanish where $k=0$, and the shifted target vanishes where $k=1$. For $d=1$ this is excluded by Lemma~\ref{l1:lem:ordinarypair} with the second product zero. For $d=0$, $A=V(\langle P_0,Z_1,l\wedge b\rangle)$, which has a good direction by Lemma~\ref{toolkit:common}, since only $l\wedge b$ can have nonzero cross part. Lemma~\ref{l4:lem:deriv} excludes the same identity. Consequently every constant external difference of $k$ is zero.

\emph{Independent external projections.}
Let $e,f$ be the two external unit directions. Here $D_fD_ec=1$. If $\tau=1$, replace $R$ by $R+\sigma k$ so that $R=z_2+v'$ with $v'\in A_0$. This preserves $kR=R$ by $k^2=k$. At $x$ put $a=D_e k$, $b'=D_fk$, $E=D_eR$, $F=D_fR$. The values of $k$ on the square are $k,k+a,k+b',k+a+b'+1$, and those of $R$ are $R,R+E,R+F,R+E+F$. Summing $(1+k)R=0$ on the square gives
\[
R=E+F+aF+b'E.
\]
Each of $a,b',E,F$ lies in $A_0$. Every product from $A_0$ has degree at most three, since $P_0\wedge Z_1=0$ and the exterior square of each generator is zero in characteristic two. Thus the right side has degree at most three, whereas $R$ has carry coefficient one.

If $\tau=0,\sigma=1$, both external differences of $k$ are constant and hence zero. Taking the mixed difference of $(1+k)R=0$ then gives $(1+k)D_fD_eR=1+k=0$. The case $\tau=\sigma=0$ is treated below.

\emph{An external direction with no generator difference.}
It remains to treat dependent external projections with arbitrary $\tau,\sigma$, and independent projections with $\tau=\sigma=0$. In the dependent case choose a nonzero external $e$ in the common kernel of $l,b$, so $D_ec=0$. In the independent case choose either external unit direction, since neither $k$ nor $R+z_2$ uses $c$. In both cases $D_e k=D_e u$ is constant and therefore zero by the first paragraph. The difference $D_eR=D_ez_2+D_ev$ is a nonconstant affine function of the low bits alone. Taking the difference of $(1+k)R=0$ shows $k=1$ on the two distinct low-bit fibers where $D_eR=1$.

Write $(s,t)=(a_0,b_0)$ and denote free linear parts by bars, in the four coordinates $(a_1,a_2,b_1,b_2)$. If $\tau=1$, the free second differences on one of those fibers give $\bar l\wedge\bar b=0$. As $\bar b\ne0$, $\bar l=\eta\bar b$. Using $b^2=b$ gives $c=C(s,t)b$ with $C$ a function of the low bits. Also $b$ has an external coefficient, since otherwise $l$ would have none and $c$ would be core supported. When $\tau=0$, take $C=0$ for the slope of $k$. In either case the free slope of $k$ on a fiber is
\[
L+\zeta(t a_1+s b_1)+C(s,t)\bar b
\]
for a fixed free linear form $L$ and the $Z_1$ coefficient $\zeta$ of $u$. Subtract the two zero slopes on the fibers where $k=1$. If $\tau=1$, an external coefficient of $b$ forces their $C$ values equal. This equality is automatic for $\tau=0$. The two distinct low-bit pairs then force $\zeta=0$. In particular $k$ is constant on each fiber if $\tau=0$. If $\tau=1$, any nonzero free slope of $k$ is $\bar b$.

On each off-ideal fiber $R$ is nonzero. Its free quadratic part is $a_1\wedge b_1+\sigma\bar l\wedge\bar b$. If this vanished, then $\sigma=1$ and the span of $\bar l,\bar b$ would equal $\langle a_1,b_1\rangle$, by the support of a nonzero wedge. This would make $c$ core supported. If $\tau=1$, the free quadratic part is exactly $a_1\wedge b_1$ and $D_iR$ is constant on the fiber for any external unit direction $i$. A nonzero free slope of $k$ would give $D_i k=1$ on the fiber for some such $i$, whence $R=kD_iR$ there by the same two evaluations. The right side is affine on the fiber. Its mixed middle-bit difference is zero, whereas that of $R$ is one. Thus $k$ is constant on every off-ideal fiber, and $kR=R$ with $R$ nonzero there forces $k=1$.
\end{proof}

\section{Four linear gates}\label{sec:lambda4}

\begin{lemma}\label{l4:lem:goodspan}
Let $P_1,P_2,P_3$ be wedges, allowing zero. If $Z_1\in Q=\langle P_0,P_1,P_2,P_3\rangle$, then $Q$ has a good direction.
\end{lemma}
\begin{proof}
Write $Z_1=c_0P_0+\sum_{i\in S}P_i$. Since $Z=Z_1+c_0P_0$ has rank four, $|S|$ is at least two. If $|S|=2$, Lemma~\ref{l4:lem:support} puts both summands in $\Lambda^2(U)$, so Lemma~\ref{toolkit:common} applies. Otherwise $P_1+P_2+P_3=Z$ and their cross parts sum to zero. If one cross part vanishes the others agree and Lemma~\ref{toolkit:common} applies. Assume all three are nonzero.

Use the types of Lemma~\ref{l4:lem:types}. The $a_2\wedge b_2$ coefficient makes the number of full forms even. With two full forms $(a_2+x_i)\wedge(b_2+y_i)$, $i=1,2$, and a type-$v$ form $(v_1a_2+v_2b_2+x_3)\wedge r$, the vanishing cross sum gives $x_2=x_1+v_2r$ and $y_2=y_1+v_1r$. Their sum is $(v_1x_1+v_2y_1+x_3)\wedge r$, contradicting the rank of $Z$.

With three type-$v$ forms $l_i\wedge r_i$, if the types all agree the external direction annihilating their common type kills $Q$. If precisely two agree, the independent types in $\sum_i v_i\otimes r_i=0$ force the remaining $r_i$ to be zero. If all are distinct, the third type is the sum of the first two, so $r_1=r_2=r_3$. Then $Z=(l_1+l_2+l_3)\wedge r_1$, again impossible. Lean checks the type alternatives by finite computation in \nolinkurl{zeroFull_finite}.
\end{proof}

\begin{theorem}\label{l4:thm:main}
Let $P_1,P_2,P_3$ be wedges, allowing zero, and let $Q=\langle P_0,P_1,P_2,P_3\rangle$ contain $Z_1$. Then \eqref{l4:eq:star} has no solution. Consequently, no circuit with five AND gates, four of them linear, computes $(z_0,z_1,z_2)$.
\end{theorem}
\begin{proof}
Lemmas~\ref{l4:lem:goodspan} and~\ref{l4:lem:deriv} give the single-product conclusion for any such wedge span, including redundant generators. Normalize a putative circuit with four linear gates. If normalization gives five linear gates, Theorem~\ref{thm:lambda5} applies. Otherwise $V_4=V(Q)$ and $z_2\notin V_4$ by degree. Thus the last gate must place $z_2$, and $z_1$ must already be available in $V_4$ since otherwise $z_1+z_2$ would lie there, also impossible by degree. This gives $Z_1\in Q$ and the excluded single-product identity.
\end{proof}

\section{Three linear gates}\label{sec:lambda3}
Let $Q_3=\langle P_0,P_1,P_2\rangle$ with $P_i=l_i\wedge r_i$ for $i=1,2$. Write $V_3=V(Q_3)$, $g_3=fh$ with $f,h\in V_3$, and $V_4=V_3+\langle g_3\rangle$.

\begin{theorem}\label{l3:thm:main}
No circuit with five AND gates, exactly three of them linear, computes $(z_0,z_1,z_2)$.
\end{theorem}

We say that $Q$ has a $U$-direction if there are $e,\ell$ with $e_{a_0}=e_{b_0}=0$, $(e_{a_1},e_{b_1})\ne(0,0)$, and $\iota_e Q\subseteq\langle\ell\rangle$, where $\ell=0$ or $\ell$ has an external coefficient.

\begin{lemma}\label{l3:lem:udir}
If $Q_3$ has no good direction, then it has a $U$-direction.
\end{lemma}
\begin{proof}
If one generator has zero cross part, Lemma~\ref{toolkit:common} gives a good direction. Two generators of the same type $v$ share an external killing direction. In the remaining cases write $e=m+n$ in the middle-bit and external direction planes, respectively, with $m\ne0$. This kills $P_0$.

If both forms are full, choose any $m\ne0$ and solve uniquely for $n$ to kill $l_1,r_1$, whose external coefficients are independent. If just one is type $v$, take it first, choose nonzero $m$ killing $r_1\in U$, and solve $l_1(m+n)=0$ for $n$. In both cases $P_1$ is killed and $\ell=\iota_eP_2=r_2(e)l_2+l_2(e)r_2$ either is zero or has an external coefficient, since $P_2$ is full.

If the types differ and neither is full, again choose nonzero $m$ killing $r_1$, and solve for $n$ killing $l_1$ and $l_2$ on $m+n$. Their external coefficients are independent, so the two equations are solvable. Now $\iota_eP_2=r_2(e)l_2$, which has the required property. Lean checks the direction choices by finite computation in \nolinkurl{kill_v_fin} and \nolinkurl{kill_vv_fin}.
\end{proof}

Here is the common hyperplane argument for $z_1$. Suppose $\iota_e Q\subseteq\langle\ell\rangle$ but $\iota_e Z_1\notin\langle\ell\rangle$. If $fh+z_1\in V(Q)$ with $f,h\in V(Q)$, normalize $D_e f$ to a constant as in Lemma~\ref{l4:lem:deriv}. Its proof applies with $A=D_e(fh)$, whose linear part is outside $\langle\ell\rangle$ by the quadratic derivative formula. The modified product is still quadratic and equals either $fA$ or $Ah$. It vanishes on the hyperplane $A=0$. Lemma~\ref{toolkit:w} therefore gives a wedge in $Z_1+Q$.

\begin{lemma}[Obstruction to placing $z_1$]\label{l3:lem:A}
If $Q_3$ has no good direction, no $f,h\in V_3$ satisfy $fh+z_1\in V_3$.
\end{lemma}
\begin{proof}
A $U$-direction exists by Lemma~\ref{l3:lem:udir}. Then $\iota_e Z_1=e_{b_1}a_0+e_{a_1}b_0$ is nonzero and has no external coefficient, so it is outside $\langle\ell\rangle$. The hyperplane argument gives a wedge $P_3\in Z_1+Q_3$. Lemma~\ref{l4:lem:goodspan} gives a good direction for $Q_3+\langle P_3\rangle$ and hence for $Q_3$, a contradiction.
\end{proof}

\begin{lemma}[Third difference along a $U$-direction]\label{l3:lem:third}
If $Q$ has a $U$-direction, then for each $\varepsilon\in\F$ no $f,h\in V(Q)$ satisfy $fh+z_2+\varepsilon z_1\in V(Q)$.
\end{lemma}
\begin{proof}
Suppose $g=fh+z_2+\varepsilon z_1\in V(Q)$. On $\ker\ell$, $D_ef$ and $D_eh$ are constants $c,d$. Therefore on that hyperplane
\[
D_ez_2=ch+df+cd+D_eg+\varepsilon D_ez_1.
\]
The right side has degree at most two. Choose an external unit vector $w$ with $\ell(w)=1$ if $\ell\ne0$, or $w=0$ otherwise. For a core coordinate $i$ set $u_i=e_i+\ell(e_i)w$, so $u_i\in\ker\ell$ and has core projection $e_i$. The third difference on any three such edges is zero on the right, as three differences annihilate a quadratic. The only cubic terms on the left are
\[
a_0b_0(e_{a_1}b_1+e_{b_1}a_1).
\]
All other terms have degree at most two, by expansion of $D_ez_2$ with $e_{a_0}=e_{b_0}=0$. If $e_{a_1}=1$, take the cube at zero on $u_{a_0},u_{b_0},u_{b_1}$, on which the third difference of the displayed cubic is one. Otherwise $e_{b_1}=1$ and the cube on $u_{a_0},u_{a_1},u_{b_0}$ gives one. This contradicts the right side. Lean checks the two target cube sums in \nolinkurl{T3_Dz2_a} and \nolinkurl{T3_Dz2_b}.
\end{proof}

\begin{lemma}[Obstruction to placing $z_2$ or $z_2+z_1$]\label{l3:lem:C}
For $\varepsilon\in\F$, no $f,h\in V_3$ satisfy $fh+z_2+\varepsilon z_1\in V_3$.
\end{lemma}
\begin{proof}
If $Q_3$ has a good direction, adjoining $Z_1$ preserves it, and the asserted placement would give $fh+z_2\in V(Q_3+\langle Z_1\rangle)$, contrary to Lemma~\ref{l4:lem:deriv}. Otherwise use Lemmas~\ref{l3:lem:udir} and~\ref{l3:lem:third}.
\end{proof}

\begin{proposition}[Last gate after placement of $z_1$]\label{l3:prop:B}
If $Z_1\notin Q_3$ and $g_3+z_1\in V_3$, then $z_2\notin V_4+\langle FH\rangle$ for $F,H\in V_4$.
\end{proposition}
\begin{proof}
Lemma~\ref{l3:lem:A} forces a good direction for $Q_3$. The placement gives $V_4=V_3+\langle z_1\rangle=V(Q_3+\langle Z_1\rangle)$, by subtracting a multiple of $z_1$ from the quadratic part. This space inherits the good direction. Degree excludes $z_2\in V_4$, and Lemma~\ref{l4:lem:deriv} excludes $FH+z_2\in V_4$.
\end{proof}

\begin{lemma}\label{l3:lem:B3U}
If $Z_1\in Q_3$, then $Q_3\subseteq\Lambda^2(U)$.
\end{lemma}
\begin{proof}
The representation of $Z_1$ must use both $P_1,P_2$ since $Z_1+cP_0$ has rank four for either $c$. Apply Lemma~\ref{l4:lem:support}.
\end{proof}

\begin{proof}[Proof of Theorem~\ref{l3:thm:main}]
Normalize the circuit. If the number of linear gates increases to four, Theorem~\ref{l4:thm:main} applies. In the normalized three-linear-gate case, degree and Lemma~\ref{l3:lem:C} exclude both $z_2$ and $z_2+z_1$ from $V_4$. The final extension has dimension at most one, so $z_1\in V_4$. Otherwise the two target classes would coincide there and their sum would already be available. If $Z_1\notin Q_3$, $g_3$ places $z_1$, and Proposition~\ref{l3:prop:B} excludes the last gate. If $Z_1\in Q_3$, Lemma~\ref{l3:lem:B3U} puts $Q_3$ inside $\Lambda^2(U)$, and Lemma~\ref{toolkit:coset} excludes the last gate on $V_3+\langle g_3\rangle$.
\end{proof}

\section{Two linear gates}\label{sec:lambda2}
Throughout this section let $P=l\wedge r$, $Q=\langle P_0,P\rangle$, and $S=Q+\langle Z_1\rangle$.

\begin{theorem}\label{l2:thm:main}
No normalized five-AND circuit with exactly two linear gates computes $(z_0,z_1,z_2)$.
\end{theorem}

\begin{lemma}\label{l2:lem:early}
If $\Phi(P)\ne0$, then no $f,h\in V(Q)$ satisfy $fh+z_1\in V(Q)$.
\end{lemma}
\begin{proof}
Choose the factors so that $l$ has nonzero external projection. Choose a nonzero middle-bit direction $m$ killing $r$ if $r$ has zero external projection. More generally, if the projections are dependent, replace $r$ by $r+l$ when they agree, which leaves $P$ unchanged, and make that same choice of $m$. Solve for an external component $n$ such that $l(m+n)=0$. If the projections are independent, take any nonzero $m$ and solve both $l(m+n)=r(m+n)=0$. In each case $e=m+n$ kills $Q$, and $\iota_eZ_1=e_{b_1}a_0+e_{a_1}b_0\ne0$. Lean checks the direction choice by finite computation in \nolinkurl{kill_two_fin}.

The hyperplane argument preceding Lemma~\ref{l3:lem:A} gives a wedge in $Z_1+Q$. It must have the form $Z_1+\alpha P_0+P$, since $Z_1+\alpha P_0$ has rank four. Lemma~\ref{l4:lem:support} would then give $\Phi(P)=0$.
\end{proof}

\begin{lemma}\label{l2:lem:onebase}
For $f,h\in V(S)$, one has $z_2\notin V(S)+\langle fh\rangle$.
\end{lemma}
\begin{proof}
The generators of $Q$ have at most one distinct nonzero cross part. Lemma~\ref{toolkit:common} gives a good direction, which survives adjoining $Z_1$. Degree excludes $z_2\in V(S)$ and Lemma~\ref{l4:lem:deriv} excludes $fh+z_2\in V(S)$.
\end{proof}

\clearpage
\begin{lemma}[Two ordinary products on a cube]\label{l2:lem:cube}
If $\Phi(P)\ne0$, then $FH+fh+z_2\notin V(S)$ for all $F,H,f,h\in V(S)$.
\end{lemma}
\begin{proof}
Swap $l,r$ if needed so that $l(w)=1$ for some external unit direction $w$. Use the 16-point cube consisting of $x=u+l(u)w$, where $u$ runs over the core direction space with coordinates $(a_0,a_1,b_0,b_1)=(a,b,p,q)$. This cube is in $\ker l$ and projects bijectively onto the core coordinates. All elements of $V(S)$ are affine plus a combination of $a_0b_0,lr,z_1$. On the cube their restrictions are linear combinations of
\[
1,\ a,\ b,\ p,\ q,\ ap,\ aq+bp.
\]
All products of two such restrictions have degree at most three. The only products of quadratic generators to inspect are their two squares and $ap(aq+bp)=apq+abp$. Their parity is therefore zero, by the even multiplicity of every monomial of degree at most three on four bits. The target has parity one: its quadratic part stays quadratic under the cube parametrization, and its carry is $abpq$. Taking the parity of a proposed identity gives a contradiction. Lean checks the product sum by finite computation in \nolinkurl{csum_phi_basis} and bilinearity in \nolinkurl{csum_phi_mul}, and checks the target sum in \nolinkurl{csum_z2}.
\end{proof}

\begin{proposition}[One signal]\label{l2:prop:onehigh}
For $f,h\in V(Q)$ and $g=fh$, put $W=V(S)+\langle g\rangle$. Then $z_2\notin W$, and no product of two elements of $W$ belongs to $z_2+W$.
\end{proposition}
\begin{proof}
If $\Phi(P)=0$, $S\subseteq\Lambda^2(U)$ and Lemma~\ref{toolkit:coset} applies. Otherwise Lemma~\ref{l2:lem:onebase} excludes $z_2\in W$. For a proposed product, apply Lemma~\ref{toolkit:feedback}. Its ordinary alternative is $FH+z_2+\sigma g\in V(S)$ with $F,H\in V(S)$. Lemma~\ref{l2:lem:onebase} handles $\sigma=0$ and Lemma~\ref{l2:lem:cube} handles $\sigma=1$.

In the feedback alternative put $R=k(g+F)$ with $k,F\in V(S)$ and $v=R+z_2\in V(S)$. Then $kR=R$. Take $d=0$ and $c=lr$ in Lemma~\ref{l2:lem:lastfactor}, swapping the factors so the right factor has an external coefficient. The nonzero cross part ensures $c$ is noncore. This gives $k\equiv1\pmod I$. The congruence and $z_2\equiv q_2\pmod I$ imply $fh\equiv q_2+F+v\pmod I$, excluded by Lemma~\ref{toolkit:modfeedback}. As a corollary of the rigidity conclusion, $k=1$ or $k=1+P_0$, since $k+1$ is quadratic and lies in $I$ (see the observation after Lemma~\ref{l2:lem:ideal}).
\end{proof}

\paragraph{Three-product tail.}
Let $P=l\wedge r$ for arbitrary homogeneous linear forms $l,r$,
and let $B=V(\langle P_0,P\rangle)$. Starting with $B$, form
$B_{i+1}=B_i+\langle f_i h_i\rangle$ for $i=0,1,2$, where
$B_0=B$ and $f_i,h_i\in B_i$. Then $z_1,z_2$ cannot both belong
to $B_3$. The proof below establishes this assertion. The normalized
two-linear-gate case is its instance $B_i=V_{2+i}$.

\begin{proof}[Proof of the tail assertion and Theorem~\ref{l2:thm:main}]
Suppose both targets belong to $B_3$ and write $p_i=f_ih_i$. Since $Z_1$ and $Z_1+P_0$ have rank four and $P$ has rank at most two, $z_1\notin B_0$. If $z_1\in B_1$, then $p_0+z_1\in B_0$. Lemma~\ref{l2:lem:early} forces $\Phi(P)=0$, and $B_1\subseteq V(S)$. Apply Lemma~\ref{toolkit:coset} to the core base $V(S)$ with signal $p_1$, a product from $B_1$, to exclude $z_2$ from $B_3$.

If $z_1\notin B_1$, set $W=V(S)+\langle p_0\rangle$. Proposition~\ref{l2:prop:onehigh} excludes $z_2$ both in $W$ and after adjoining the product $p_1$ from $B_1\subseteq W$. The same holds for $z_2+z_1$ because $z_1\in W$. Thus neither lies in $B_2$. The final extension has dimension at most one, so $z_1\in B_2$. Then $p_1+z_1\in B_1$ and $B_2\subseteq W$. Applying the proposition to $p_2$ excludes $z_2$ from $B_3$.
\end{proof}

\section{One linear gate}\label{sec:lambda1}
Here $g_0=a_0b_0$, $V_1=\Aff+\langle a_0b_0\rangle$, and $g_1,g_2,g_3,g_4$ are nonlinear.

\begin{theorem}\label{l1:thm:main}
No normalized five-AND circuit with exactly one linear gate computes $(z_0,z_1,z_2)$.
\end{theorem}
We normalize the first nonlinear gate and apply the three-product tail argument.

\begin{lemma}[Cubic first-gate normal form]\label{l1:lem:cubic}
Every product $g_1$ of two factors in $V_1$, at least one of them nonaffine, has a representative
\[
g_1\equiv(P_0+A)B\pmod{V_1},
\]
with $A,B$ homogeneous linear functions. The original product has degree three exactly when $B\notin\langle a_0,b_0\rangle$. Its cubic part is $P_0\wedge B$.
\end{lemma}
\begin{proof}
If the factors are dependent modulo constants, their product lies in $V_1$ and the representative with $B=0$ suffices. Otherwise the factor plane has a nonzero quadratic coefficient, and the following operations apply.
The quadratic coefficients of the factors form a nonzero vector in the one-dimensional space $\langle P_0\rangle$. By exchanging factors or replacing one by their sum, assume one factor has quadratic part $P_0$ and the other is affine. Such plane operations change the product only by an available factor. Remove factor constants, also changing the product only by elements of $V_1$. We obtain $(P_0+A)B$.

The affine factor is a nonzero homogeneous linear function $B$. The cubic part is $P_0\wedge B$, which vanishes exactly for $B\in\langle a_0,b_0\rangle$.
\end{proof}

\begin{lemma}[Quadratic first gate]\label{l1:lem:reduce}
If $\deg g_1\le2$ and $g_1+(P_0+l)b\in V_1$ for homogeneous $l,b$, then $V_1+\langle g_1\rangle\subseteq V(\langle P_0,l\wedge b\rangle)$.
\end{lemma}
\begin{proof}
The cubic coefficient test gives $b\in\langle a_0,b_0\rangle$, so $P_0b$ is a scalar multiple of $P_0$. Hence $g_1\equiv lb\pmod{V_1}$, which proves the inclusion.
\end{proof}

\begin{lemma}[The first nonlinear gate places no remaining output]\label{l1:lem:first}
For every $g_1=fh$, $f,h\in V_1$, none of $z_1,z_2,z_1+z_2$ belongs to $V_2=V_1+\langle g_1\rangle$.
\end{lemma}
\begin{proof}
Set $Q_*=\langle P_0,a_2\wedge b_2\rangle$.
We have $V_1\subseteq V(Q_*)$, and $z_1\notin V_1$
since $Z_1$ has rank four. If $z_1\in V_1+\langle fh\rangle$,
then $fh+z_1\in V_1\subseteq V(Q_*)$.
The extra generator of $Q_*$ has nonzero cross part, so
Lemma~\ref{l2:lem:early} excludes this identity.
If both factors are affine, their product has degree at most two.
Otherwise Lemma~\ref{l1:lem:cubic} gives a representative of degree
at most three modulo $V_1$, so the product itself has degree at most
three. Hence neither $z_2$ nor $z_1+z_2$ lies in the extended space.
\end{proof}

\begin{lemma}[Early feedback]\label{l1:lem:earlyfeedback}
Let $c=(P_0+l)b$ with $b\notin\langle a_0,b_0\rangle$. For $k,F\in V_1$, one has $k(c+F)+z_1\notin V_1$.
\end{lemma}
\begin{proof}
Suppose $R=k(c+F)=z_1+v$ with $v\in V_1$. Then $kR=R$. Write $s=a_0,t=b_0$. On each low-bit fiber, $R$ is affine in the free coordinates, with middle-bit slopes $(v_{a_1}+t,v_{b_1}+s)$. Here the coefficients of $v$ in the free variables are fixed across the fibers.

If $k$ had a nonzero free slope, then along that unit direction $i$ we would have $D_i k=1$ on every fiber. The two evaluations of $kR=R$ give $R=kD_iR$. Since $D_iR$ is constant on each fiber, the middle-bit slope of $R$ would be either zero or the fixed middle-bit slope of $k$. But its displayed slopes range over all four pairs. Thus $k$ is constant on each fiber. At least one fiber has $k=1$, since $R$ has a nonzero slope somewhere.

On a fiber with $k=1$, $c+F=R$ is affine in the free coordinates, so the free quadratic part $\bar l\wedge\bar b$ vanishes. With $\bar b\ne0$ this yields $\bar l=\eta\bar b$. Consequently on every fiber $c$ is a scalar multiple of $b$, since $b^2=b$. Thus the middle-bit slope of $R$ is zero on each zero fiber of $k$, and on the others belongs to the affine line of slopes $F_{\rm mid}+\langle b_{\rm mid}\rangle$. These possibilities cover at most three pairs, whereas all four occur.
\end{proof}

\begin{lemma}[Early $z_1$ forces a core cubic signal]\label{l1:lem:earlycore}
Suppose $g_1$ is cubic and $z_1\in V_2+\langle fh\rangle$ for $f,h\in V_2$. Then $g_1$ is core supported.
\end{lemma}
\begin{proof}
Use $g_1\equiv(P_0+A)B\pmod{V_1}$ with $A,B$ homogeneous linear. Apply Lemma~\ref{toolkit:feedback} to the placement in $V_1+\langle g_1\rangle$. Lemma~\ref{l1:lem:earlyfeedback} excludes its feedback alternative, so there is a product of two factors from $V_1$ placing $z_1$ modulo $V_1+\langle g_1\rangle$. By Lemma~\ref{l1:lem:first} applied to that product, its output correction must include $g_1$. The product is therefore cubic. Normalize it to $(P_0+C)D$. Cancellation of cubic terms gives $L=B+D\in\langle a_0,b_0\rangle$. With $M=A+C$, comparison of quadratic parts gives
\[
Z_1+\epsilon P_0=M\wedge B+C\wedge L.
\]
The left side has rank four and support $U$. Both summands must be nonzero rank-two forms, and Lemma~\ref{l4:lem:support} puts their supports in $U$. Hence $C,L,M,B\in U$, and therefore $A,B\in U$. The normalized $g_1$ is core supported; restoring elements of $V_1$ preserves that property.

\end{proof}

\subsection{The noncore branch and the theorem}
\begin{lemma}[Noncore intermediate-space obstruction]\label{l1:lem:noncore}
Let $c=(P_0+l)b$ be noncore with $l,b$ homogeneous and $b\notin\langle a_0,b_0\rangle$. Put $B=\Aff+\langle P_0,c\rangle$, $A=B+\langle Z_1\rangle$, and $g=fh$ a product with $f,h\in B$. Put $W=A+\langle g\rangle$. Then $z_2\notin W$, and $FH+z_2\notin W$ for all $F,H\in W$.
\end{lemma}
\begin{proof}
Lemma~\ref{l1:lem:ordinarypair} excludes $z_2\in A+\langle g\rangle$ and any further product with both factors in $A$, including a $g$ output correction: these are sums of at most two products from $A$.

Otherwise Lemma~\ref{toolkit:feedback} gives $k(g+F)=z_2+v$ with $k,F,v\in A$. With $R=k(g+F)$ one has $kR=R$. Lemma~\ref{l2:lem:lastfactor} gives $k\equiv1\pmod I$. Use $c\equiv lb\pmod I$ and put $Q=\langle P_0,l\wedge b\rangle$.

Replacing $c$ by $lb$ in the factors of $g$ gives $f',h'\in V(Q)$ with $g\equiv f'h'\pmod I$. Replacing it likewise in $F,v$ gives $F',v'\in V(Q+\langle Z_1\rangle)$. The carry vanishes modulo $I$, so
\[
f'h'\equiv q_2+v'+F'\pmod I.
\]
Lemma~\ref{toolkit:modfeedback} excludes this congruence.
\end{proof}

\begin{proof}[Proof of Theorem~\ref{l1:thm:main}]
Use the representative $c=(P_0+l)b$ of $g_1$ supplied by Lemma~\ref{l1:lem:cubic}. If $\deg g_1\le2$, Lemma~\ref{l1:lem:reduce} embeds $V_2$ in $V(\langle P_0,l\wedge b\rangle)$. Enlarge the initial space of the three remaining products to this quadratic space and keep their factors as functions. Apply the three-product tail assertion of Section~\ref{sec:lambda2} to this enlarged base.
It allows arbitrary $l,b$, including $l\wedge b=0$ and
$l\wedge b=P_0$. The three factor pairs and both target memberships
remain valid under the successive inclusions of the old wire spaces
in the enlarged spaces.

For this cubic case write $B=V_2=V_1+\langle c\rangle$ for the base space, and put $A=B+\langle z_1\rangle$. By Lemma~\ref{l1:lem:first}, $z_1\notin B$. If $z_1\in V_3$, then $c$ is core supported by Lemma~\ref{l1:lem:earlycore}. Also $g_2+z_1\in B$, hence $V_3=A$ and $g_3$ is a product from $A$. The core-base verification in the proof of Lemma~\ref{l1:lem:ordinarypair} gives the hypotheses of Lemma~\ref{toolkit:coset} with base $A$ and signal $g_3$, excluding $z_2$ after the final gate.

If $z_1\notin V_3$, put $W=A+\langle g_2\rangle$. For a core cubic use the core-base verification in the proof of Lemma~\ref{l1:lem:ordinarypair} and apply Lemma~\ref{toolkit:coset}, since $g_2$ is a product from $B\subseteq A$. For a noncore cubic use Lemma~\ref{l1:lem:noncore}. In either case $z_2$ cannot be in $W$ or placed by a product from $W$. Because $z_1\in W$, the same is true of $z_2+z_1$. Thus neither is available after the next product $g_3$ from $V_3\subseteq W$. The final extension forces $z_1\in V_4$, and $z_1\notin V_3$ now gives $g_3\in z_1+V_3$ and $V_4\subseteq W$. This excludes the final placement of $z_2$.
\end{proof}

A relaxed four-gate circuit computes $(z_1,q_2)$ on the 48 points $a_0b_0=0$. An explicit witness, re-verified on all $48$ points by an independent script, is the file \nolinkurl{verification/lambda2-lambda1-checks/quotient_relaxed_witness.json} at the release commit. The proof above uses identities modulo $I$ only after the full-domain multiplier lemma has forced the multiplier to be one there; the relaxed circuit illustrates the need for the full-domain identity.

\begin{proof}[Proof of Theorem~\ref{thm:MC}]
The circuit in Section~\ref{sec:upper} uses six AND gates. If a circuit used at most five, appending redundant affine-input AND gates would give a five-gate circuit computing the same targets. Lemma~\ref{l4:lem:normal} puts it in one of the five cases excluded by Theorems~\ref{thm:lambda5}, \ref{l4:thm:main}, \ref{l3:thm:main}, \ref{l2:thm:main}, and~\ref{l1:thm:main}. Thus $\mathrm{MC}(M_3)\ge6$.
\end{proof}

\appendix
\section{Computational verification}\label{sec:verification}
Python scripts in the repository check the finite claims used in the written proofs, each with a positive control. These scripts, the SAT certificates and the Lean formalization are supporting evidence; the written proof is self-contained.

The repository is \url{https://github.com/bishara74/lowhalf-mul3}.
The release tag is
\href{https://github.com/bishara74/lowhalf-mul3/tree/lowhalf-arxiv-v2}{\texttt{lowhalf-arxiv-v2}}, at commit
\begin{center}\small\texttt{\ReleaseCommit}\end{center}
Its \href{https://github.com/bishara74/lowhalf-mul3/tree/lowhalf-arxiv-v2/verification}{verification directory} contains the checks for each case, the checks supporting the Lean formalization, and the SAT and LRAT certificate artifacts.

\subsection{Independent certificate searches}
The first author's SAT searches ruled out five AND gates, and the second author's searches confirmed this independently. The accompanying release documents a separate 651-cube LRAT certificate~\cite{cert} based on an exact-multiplicative-complexity encoding of the kind described by Soeken~\cite{soeken}. Readers can replay the archived CNFs and LRAT proofs for $k=0,\ldots,4$ and \texttt{cover5} with \texttt{cake\_lpr}. The archive also supplies \texttt{cubes.txt}, the manifest and the generation and checker logs. The 651 per-cube proof files are absent from the archive. The \href{https://github.com/bishara74/lowhalf-mul3/blob/lowhalf-arxiv-v2/verification/certificates/REGENERATE.md}{regeneration instructions} give the commands to reconstruct and check each of them. The low-half multipliers on one, two, and three input bits have multiplicative complexities $1,3,6$. The value $6$ is Theorem~\ref{thm:MC}; the values $1$ and $3$ are established in \texttt{verification/small-sizes/} by explicit circuits and exhaustive searches showing that one gate fewer is impossible.

\section{Lean formalization}\label{sec:lean}
\LeanReleaseStatus{}
The formalization uses Lean~4 \texttt{v4.35.0-rc3}, a release candidate, and
Mathlib. Its source and axiom audit are in the
\href{https://github.com/bishara74/lowhalf-mul3/tree/lowhalf-arxiv-v2}{tagged repository}.
The formal coset lemma is stated for an abstract \texttt{CosetBase}.
The space contains the affine functions, its derivatives along $a_2$ and $b_2$
are constant, and $1,a_0,b_0$ lie in the radical of the 16-point carry pairing.
The hypotheses of Lemma~\ref{toolkit:coset} give these conditions directly.
The statement \nolinkurl{mc_lowMul3} in the self-contained challenge module
\nolinkurl{LowHalfMul3Challenge} is proved in \nolinkurl{LowHalfMul3Solution}.
The Palomar registration PALOMAR-2026-11-09-000020 (v1)~\cite{palomar}
verified commit \texttt{b6c00ef8d94ff578585b428270df1af271a96bd6}.
All \texttt{.lean} files, \texttt{lakefile.toml}, \texttt{lake-manifest.json},
\texttt{lean-toolchain}, and \texttt{comparator.json} are byte-identical
at that commit and at the release tag \texttt{lowhalf-arxiv-v2}.
The registered run at that commit builds the Challenge and Solution modules and runs
\texttt{lake comparator}. The Lean kernel, \texttt{nanoda} and \texttt{con-ron}
accept the solution, and the only axioms are \texttt{propext},
\texttt{Classical.choice} and \texttt{Quot.sound}.

\subsection{Correspondence with the Lean source}
The Lean declarations in the table are in the following namespace.
\begin{center}
\texttt{UnrestrictedBooleanMul.LowHalfMul3}
\end{center}
The names omit the subnamespace \nolinkurl{LambdaOne}. The challenge theorem \nolinkurl{mc_lowMul3}
is proved in \nolinkurl{LowHalfMul3Solution}. The support lemma is used at the
two instances formalized as \nolinkurl{lemma3}. Where Lean checks a finite step,
the manuscript gives a hand proof. The table also identifies generalizations
and differences in proof structure.

{\small
\begin{longtable}{@{}>{\raggedright\arraybackslash}p{.21\linewidth}
>{\raggedright\arraybackslash}p{.42\linewidth}
>{\raggedright\arraybackslash}p{.31\linewidth}@{}}
\toprule
Paper label & Lean declaration & Route \\
\midrule
\endfirsthead
\toprule
Paper label & Lean declaration & Route \\
\midrule
\endhead
\bottomrule
\endfoot
\nolinkurl{thm:MC} & \nolinkurl{mc_lowMul3}, \nolinkurl{mc_eq_six_of_lambda_one_two} & Padding, normalization and five exclusions \\
\nolinkurl{sec:preliminaries} & \nolinkurl{lowMul3_eval} & Finite evaluations in Lean, column carries here \\
\nolinkurl{sec:upper} & \nolinkurl{upper} & Six gates with the same outputs \\
\nolinkurl{l4:lem:normal} & \nolinkurl{normal_form_five} & Same swaps, with first factors recorded in Lean \\
\nolinkurl{l4:lem:Qonly} & \nolinkurl{wireSpace_eq_V4} & Subtracts a quadratic representative \\
\nolinkurl{l4:lem:support} & \nolinkurl{lemma3} & Lean checks two target splittings finitely \\
\nolinkurl{l4:lem:cross} & \nolinkurl{phi_eq_zero_iff} & Contractions record the cross part \\
\nolinkurl{l4:lem:noq2} & \nolinkurl{lemma6} & Lean checks the wedge exclusion finitely \\
\nolinkurl{l4:lem:types} & \nolinkurl{types} & Lean changes basis even for the zero wedge \\
\nolinkurl{thm:lambda5} & \nolinkurl{lambdaCase_five} & Degree obstruction for any gate count \\
\nolinkurl{toolkit:common} & \nolinkurl{goodDir_caseA} & Lean uses four indexed generator slots \\
\nolinkurl{l4:lem:deriv} & \nolinkurl{starFree_of_goodDir} & Lean checks the affine zero set finitely \\
\nolinkurl{toolkit:w} & \nolinkurl{pi2_wedge_of_vanish} & Tangent contractions in coordinates in Lean \\
\nolinkurl{toolkit:feedback} & \nolinkurl{feedbackNormal} & Four scalar cases with output corrections \\
\nolinkurl{toolkit:coset} & \nolinkurl{coset_obstruction}, \nolinkurl{coset_z2_not_mem} & Lean takes differences before forming the Gram matrix \\
\nolinkurl{toolkit:gram} & \nolinkurl{gram_apply} & Finite Gram identity in Lean, kernel basis here \\
\nolinkurl{l2:lem:ideal} & \nolinkurl{idealExtract} & Exceptional-fiber correction and wedge extraction \\
\nolinkurl{l2:lem:rankext} & \nolinkurl{q2NotTwoWedges} & Lean checks the trivial radical finitely \\
\nolinkurl{toolkit:modfeedback} & \nolinkurl{feedbackNoPlaceModLow} & Extraction followed by two coefficient cases \\
\nolinkurl{l1:lem:ordinarypair} & \nolinkurl{ordinary_pair_noPlace} & Pointwise slice equations in Lean \\
\nolinkurl{l2:lem:lastfactor} & \nolinkurl{lastFactorLowIdeal} & Lean states the independent $(0,1)$ result modulo $I$ \\
\nolinkurl{l4:lem:goodspan} & \nolinkurl{goodDir_main} & Lean checks the nonfull type case finitely \\
\nolinkurl{l4:thm:main} & \nolinkurl{starFree}, \nolinkurl{lambdaCase_four} & Lean starts from a normalized circuit \\
\nolinkurl{l3:lem:udir} & \nolinkurl{uDir_of_bad} & Lean checks direction choices finitely \\
\nolinkurl{l3:lem:A} & \nolinkurl{noPlace_z1_bad} & Adjoins an extracted wedge to get a good direction \\
\nolinkurl{l3:lem:third} & \nolinkurl{noPlace_z2_of_uDir} & Lean checks the tangent cube values finitely \\
\nolinkurl{l3:lem:C} & \nolinkurl{noPlace_of_goodDir}, \nolinkurl{noPlace_z2_bad}, \nolinkurl{noPlace_z21_bad} & Good direction or bad-direction obstructions for both targets \\
\nolinkurl{l3:prop:B} & \nolinkurl{lambdaCase_three_of} & Lean needs only inclusion in the enlarged space \\
\nolinkurl{l3:lem:B3U} & \nolinkurl{phi_of_Z1_mem} & Lean uses target-specific support checks \\
\nolinkurl{l3:thm:main} & \nolinkurl{lambdaCase_three} & Lean splits on the coefficient of the next product \\
\nolinkurl{l2:lem:early} & \nolinkurl{noPlace_z1_two} & Lean checks the killing direction finitely \\
\nolinkurl{l2:lem:onebase} & \nolinkurl{z2_not_mem_S_sup} & Lean calls the zero-correction case in the tail \\
\nolinkurl{l2:lem:cube} & \nolinkurl{formA_one_of_phi} & Lean checks the tilted cube sums finitely \\
\nolinkurl{l2:prop:onehigh} & \nolinkurl{oneHigh_of_phi}, \nolinkurl{twoClaim}, \nolinkurl{formB} & Lean separates the multiplier corollary \\
\nolinkurl{l2:thm:main} & \nolinkurl{lambdaCase_two}, \nolinkurl{lambdaCase_two_of} & General tail with a normalized wrapper \\
\nolinkurl{l1:lem:cubic} & \nolinkurl{first_normal_form}, \nolinkurl{first_cubic_iff} & Lean handles dependence in the coefficient cases \\
\nolinkurl{l1:lem:reduce} & \nolinkurl{quadratic_first_le} & Quadratic inclusion for arbitrary wedges \\
\nolinkurl{l1:lem:first} & \nolinkurl{no_z1_in_first}, \nolinkurl{first_degree} & Enlarges by $a_2\wedge b_2$ to exclude $z_1$ \\
\nolinkurl{l1:lem:earlyfeedback} & \nolinkurl{early_z1_feedback} & Lean checks the fiber slope subclaims finitely \\
\nolinkurl{l1:lem:earlycore} & \nolinkurl{early_z1_core} & Lean reads core support from the cross parts \\
\nolinkurl{l1:lem:noncore} & \nolinkurl{noncore_noNextZ2} & Lean substitutes the cubic signal modulo $I$ \\
\nolinkurl{l1:thm:main} & \nolinkurl{lambdaCase_one}, \nolinkurl{lambdaCase_one_of}, \nolinkurl{core_noNextZ2} & Lean shares the two tail reductions \\
\end{longtable}}

\section*{Acknowledgements and AI disclosure}
This work was developed with AI assistance, in line with the ELTE Faculty of
Informatics policy on the use of AI tools. Anthropic Claude Science, with Claude
Opus~5.5 (high effort) and Claude Sonnet~5.5 (medium effort), was used to draft the
written proofs of the four- and three-linear-gate cases, to write verification
scripts, and to write most of the Lean formalization. OpenAI Codex with GPT-6.1 Sol
(high effort) was used to draft the written proofs of the two- and one-linear-gate
cases, to write the remaining Lean proofs and further verification scripts, to merge
and revise the manuscript, and to run internal audits of it. Anthropic Claude
Opus~5.5 (medium effort) was used in a separate chat for planning, proof review and
editing. The second author used Claude Opus~5.5 (high effort) for refereeing. All AI
output was checked before it was adopted, and the Lean kernel checks the formal
proof. The authors reviewed all definitions, statements, proofs and text, and take
full responsibility for the work. No AI system is an author.

\begingroup\raggedright
\bibliographystyle{unsrt}
\bibliography{references}
\endgroup
\end{document}